\documentstyle[12pt]{article}

\newcommand{\be}{\begin{equation}}
\newcommand{\ee}{\end{equation}}
\newcommand{\bn}{\begin{eqnarray}}
\newcommand{\en}{\end{eqnarray}}

\begin{document}

\begin{center}

\noindent{\large\bf
{\huge On the dual equivalence of the self-dual and topologically massive $B\wedge F$ models coupled to dynamical fermionic matter}
}
\vspace{3mm}

\noindent{
{\Large R. Menezes$^1$, J. R. S. Nascimento$^1$, R. F. Ribeiro$^1$, and C. Wotzasek$^{2}$}
}\vspace{3mm}

\noindent

\begin{large}
{\it $^1$Departamento de F\' \i sica, Universidade Federal
da Para\'\i ba\\ 58051-970 Jo\~ao Pessoa, Para\'\i ba, Brasil. \\
$^2$Instituto de F\'\i sica, Universidade Federal do Rio de Janeiro\\
21945-970, Rio de Janeiro, Brazil. } \\
\end{large}
\end{center}

\begin{abstract}
\noindent We study the equivalence between the $B\wedge F$ self-dual ($SD_{B\wedge F}$) and the
$B\wedge F$ topologically massive ($TM_{B\wedge F}$) models including the coupling to dynamical, U(1) charged
fermionic matter. This is done through an iterative procedure of gauge embedding that produces the
dual mapping. In the interactive cases, the minimal coupling adopted for both vector and tensor fields
in the self-dual representation is transformed into a non minimal magnetic like coupling in the
topologically massive representation but with the currents swapped. It is known that to establish
this equivalence a current-current interaction term is needed to render the matter sector unchanged.
We show that both terms arise naturally from the embedding procedure.  \\
\end{abstract}

\newpage

\section{Preliminaries}

This work is devoted to the study of duality symmetry in the context of the
$B \wedge F$ theory, viz., in models presenting a topological, first-order derivative coupling between forms of different ranks that is a dimensional extension of the duality between the
self-dual (SD) \cite{TPvN} and Maxwell-Chern-Simons models (MCS) \cite{DJT}, shown by Deser and
Jackiw \cite{DJ} long time ago. To
this end we investigate the existence of a constraint of self duality in the massive, non invariant model ($SD_{B\wedge F}$)
and adopt a new dynamical embedding formalism \cite{IW,AINRW}, that is alternative to
the master Lagrangian approach, to obtain the gauge invariant $B\wedge F$ model. Our study also includes the case of dynamical fermionic matter minimally coupled to the self-dual sector.

Duality is a fascinating symmetry concept allowing the connection of two opposite regimes for the same dynamics. It plays an important role in nowadays physics,
both in the original contexts of condensed matter and Maxwell electromagnetism, as well as
in the recent research of extended objects. The existence of such a
symmetry within a model has important consequences - it can
be used to derive (exact) non perturbative results since swapping opposite
regimes allows a perturbative investigation of theories
with large coupling constants.

The study of this symmetry has received renewed interest in recent research in diverse areas in
field theory such as, supersymmetric gauge theories \cite{SW},
sine-Gordon model \cite{DC}, statistical systems \cite{JLC} and, in the
context of condensed matter models, applied for instance to planar high-T$_C$
materials, Josephson junction arrays \cite{DST} and Quantum Hall
Effect \cite{MS}. In particular the duality mapping has been of great significance in order to extend the bosonization
program from two to three dimensions with important phenomenological
consequences \cite{boson}. It also plays preponderant role in the ADS/CFT correspondence \cite{maldacena}
that illustrates the holographic principle \cite{t'hooft}.

The idea of duality has also been used in recent developments of  string
theory \cite{EK}, where different vacua are shown to be related by duality\cite{SZ}.  In this context a general procedure for constructing dual models was proposed by Busher \cite{THB} and generalized by Rocek and Verlind \cite{RV} that consists in lifting the global symmetry of the tensor fields with a new gauge field, whose field strength is then constrained to zero by the use of a Lagrange multiplier.  Integrating, sequentially, the multiplier and the gauge field yields the original action while the dual action is obtained if one integrates the gauge field together with the original tensor field, keeping the Lagrange multiplier that then plays the role of dual field to the original tensor field. This line of research was used in the investigation of bosonization as duality by Burgess and Quevedo \cite{BQ} and to discuss S-duality, the relation between strong and weak couplings in gauge theories \cite{EW}.  This procedure has also been shown to be related to canonical transformations \cite{AAGL}. Recently, this line of research has been applied in the context of the topologically massive $B\wedge F$ theory, which is related to our interest here, to study its equivalence with the Stuckelberg construction of gauge invariant massive excitations \cite{SH}.

The duality we are treating in this work deals with the equivalence between models describing the same physical phenomenon involving the presence of a topological term in a four dimensional spacetime.
It is closely related to the odd-dimensional duality involving the Chern-Simons term (CST) \cite{CST}, whose paradigm is the equivalence between SD \cite{TPvN} and MCS \cite{DJT,DJ} theories in (2+1) dimensions.
As shown in \cite{DJ}, in three dimensions there are two different ways to describe the dynamics of a single, freely propagating spin one massive mode, using either the SD theory \cite{TPvN} or the MCS theory.
They also established the identification that relates the basic field of the SD model with the dual of the MCS field \cite{DJ}. This correspondence displays the way the gauge symmetry of the MCS representation, gets hidden in the SD representation \cite{DJ}.
It is well understood by now that it is the presence of the topological and gauge invariant Chern-Simons term the responsible for the essential features manifested by the three dimensional field theories, while in the four dimensional context this role is played by the $B\wedge F$ term.
To extend this duality symmetry relation and study its consequences in the context of four dimensional field theories with $B\wedge F$ term, is the main purpose of this work.

The study of gauge theories with a topological term, in (3+1) dimensions, has received considerable attention recently.  Among other possibilities the $B\wedge F$ term is interesting for providing a gauge invariant mechanism to give mass to the gauge field and to produce statistical transmutation in (3+1) dimensions.
Here $B$ is a Kalb-Ramond field, i.e., a totally antisymmetric tensor potential (a potential 2-form) while $F = dA$ is the field strength of
the one-form potential $A$.  An abelian antisymmetric tensor potential was probably first used in the context of the
particle theory to describe a massless particle of zero-helicity \cite{OP,SD}.  It reappeared later on in the context
of fundamental strings \cite{KR,RW}, has been used to study cosmic strings \cite{VV,DS,RLD} and to put topological charge (hair) on black holes\cite{MJB,ABL,BAC}. The free theory of a rank-2 antisymmetric tensor has also been intensively studied both classically \cite{RKK} and quantically \cite{BK,MO} and has been shown to be dynamically dual (under the Hodge mapping) to a massless scalar field (zero-form).

In this work we study duality in field theories involving the $B\wedge F$ term. 
To this end, in the next section, we investigate the gauge non invariant $SD_{B\wedge F}$ model, define a new, non-Hodge, (derivative) duality operation  and show the existence of self-duality.  Next, in Section III, we apply an iterative dynamical embedding procedure to construct an invariant theory out of the self-dual $B\wedge F$ model - the topologically massive $B\wedge F$ model ($TM_{B\wedge F}$).  This is a gauge embedding procedure that is done with the inclusion of counter
terms in the non invariant action, built with powers of the Euler vectors and tensors (whose kernels give the field equations for the potentials $A$ and $B$) to warrant the dynamical equivalence.
Such construction discloses hidden
gauge symmetries in such systems.
One can then consider the non-invariant
model as the gauge fixed version of a gauge theory.
A deeper and more illuminating interpretation of these systems is then obtained.
The advantage in having a gauge theory lies in the fact that the underlying gauge symmetry
allows us to establish a chain of equivalence among different models by
choosing different gauge fixing conditions. In Section IV we consider the minimal coupling with fermionic
matter.  Our results are discussed in the final section of the paper.

\section{Self Dual $B\wedge F$ Theory}

The model with a built-in SD constraint in (2+1) dimensions was proposed in \cite{TPvN} as an alternative to
the concept of topologically massive modes proposed in \cite{DJT}. The former is a non gauge
invariant, first order model, while the later is a second order gauge invariant
formulation, both making use of the topological Chern-Simons term.  In this section
we want to formulate and study a first order, non gauge invariant model, making use of the
topological $B\wedge F$ term and prove the existence of self duality property as a consequence of a built in
SD constraint.

The model in question shows the coupling of a vector field potential $A_\mu$ with a tensor field potential $B_{\mu\nu}$ \cite{ABL} as,
\begin{equation}
\label{PB10}
{\cal L}_{SD}^{(0)} = \frac 12 m^2 A_\mu A^\mu - \frac 14 B_{\mu\nu} B^{\mu\nu}
+\frac {\chi\,\theta}4 \epsilon^{\mu\nu\lambda\rho} B_{\mu\nu} F_{\lambda\rho} \; ,
\end{equation}
where the superscript index in the Lagrangean is the counter of the iterative algorithm
to be implemented in the sequel, $\chi = \pm 1$ will be shown to display the self or anti-self duality, $\theta$ is the coupling constant and the field strength of the basic potentials are,
\begin{eqnarray}
\label{PB20}
F_{\mu\nu} &=& \partial_\mu A_\nu - \partial_\nu A_\mu \nonumber\\
H_{\mu\nu\lambda} &=& \partial_\mu B_{\nu\lambda} + \partial_\nu B_{\lambda\mu} + 
\partial_\lambda B_{\mu\nu}\, .
\end{eqnarray}
The coefficients of the mass terms are so chosen to give mass dimension one and two, respectively, to the potentials
$A_\mu$ and $B_{\mu\nu}$ which, consequently, keeps dimensionless the coupling constant $\theta$ in the $B\wedge F$ term.
Here the potentials play an active role in the duality transformations.  This shall be in contrast with the dynamical matter field, to be considered latter on that, although coupled to the potentials, are passive fields (spectators) in the duality mapping.
It is immediate to work out the equations of motion of the basic potentials $A_\mu$ and $B_{\mu\nu}$ to obtain, respectively,
\begin{eqnarray}
\label{PB30}
A^\mu &=& -\, \frac {\chi\,\theta}{2 m^2} \epsilon^{\mu\nu\lambda\rho}\partial_\nu B_{\lambda\rho}\nonumber\\
B^{\mu\nu} &=&  \chi\,\theta \:\epsilon^{\mu\nu\lambda\rho}\partial_\lambda  A_{\rho}
\end{eqnarray}
satisfying the constraints
\begin{eqnarray}
\label{PB40}
\partial_\mu A^\mu &=& 0\\
\partial_\mu B^{\mu\nu} &=& 0
\end{eqnarray}
identically.  The equations (\ref{PB30}) constitute a set of first-order
coupled equations that can be combined into a decoupled second-order, massive, wave equations as
\begin{eqnarray}
\label{PB70}
\left(\partial_\alpha\partial^\alpha + \frac{m^2}{\theta^2}\right) F&=& 0 \:\:\:\:\: ; \:\:\:\:\:F = \left\{A_\mu ; B_{\mu\nu}\right\}
\end{eqnarray}
whose mass depends crucially on the value of the coupling constant.

Next, we discuss the self-duality inherent to the above theory.  To this end we define a new derivative
duality operation by means of a set of star-variables as
\begin{eqnarray}
\label{PB90}
\mbox{}^*A_\mu &\equiv & -\,  \frac {\theta}{2m^2} \epsilon_{\mu\nu\lambda\rho} \partial^\nu B^{\lambda\rho}\nonumber\\
\mbox{}^*B_{\mu\nu} &\equiv & \theta \epsilon_{\mu\nu\lambda\rho} \partial^\lambda A^{\rho} 
\end{eqnarray}
With this definition we obtain, for the double duality operation, the relations
\begin{eqnarray}
\label{PB110}
\mbox{}^*\left(\mbox{}^* F\right) =  F \:\:\:\:\: ; \:\:\:\:\: F = \left\{A_\mu ; B_{\mu\nu}\right\}
\end{eqnarray}
after use of the equations of motion (\ref{PB70}).  This is important because it validates the
notion of self (or anti-self) duality
\begin{eqnarray}
\label{PB130}
\mbox{}^* F = \chi\, F \:\:\:\:\: ; \:\:\:\:\: F = \left\{A_\mu ; B_{\mu\nu}\right\}
\end{eqnarray}
as a solution for the field equations, very much like the three-dimensional SD
model. However, this conceptualization of duality operation and self-duality {\it in four-dimensions} is new.

Before we start the iterative procedure for the transformation of the $SD_{B\wedge F}$
model into a topological $B\wedge F$ model let us digress on the consequences of the self-duality relation (\ref{PB130}).
Notice first that under the usual gauge transformations of the potentials
\begin{eqnarray}
\label{PB140}
A_\mu &\to &  A_\mu + \partial_\mu \Lambda \nonumber\\
B_{\mu\nu} &\to &  B_{\mu\nu} +\partial_\mu\Lambda_\nu -\partial_\nu\Lambda_\mu
\end{eqnarray}
the fields strengths $F_{\mu\nu}$ and $H_{\mu\nu\lambda}$ are left invariant. Therefore, although the basic potentials are gauge dependent their duals, defined in (\ref{PB90}), are not. This situation parallels the three-dimensional case involving the Chern-Simons term which is the origin for the presence of a hidden (gauge) symmetry in the SD model of \cite{TPvN} while it is explicit in the topologically massive model of \cite{DJT}. Here too the $SD_{B\wedge F}$ model hides the gauge symmetry (\ref{PB140}) that is explicit in the $TM_{B\wedge F}$ model.
Let us next consider the direct automorphism
\begin{equation}
\label{PB150}
{\cal L} (A,B) \longrightarrow \mbox{}^*{\cal L} (\mbox{}^* A, \mbox{}^* B)
\end{equation}
Since it is constructed with the dual fields it is automatically gauge invariant and reads
\begin{equation}
\label{PB160}
\mbox{}^*{\cal L} (\mbox{}^* A, \mbox{}^* B) = \frac 1{12 m^2} H_{\alpha\beta\lambda}H^{\alpha\beta\lambda} - \frac 18 F_{\alpha\beta}F^{\alpha\beta}
+ \frac 12 \epsilon^{\alpha\beta\lambda\rho} \partial^\sigma F_{\sigma\rho} H_{\alpha\beta\lambda}
\end{equation}
It is a simple algebra to check that the equations of motion for the SD action
(\ref{PB10}) are also solutions for the field equation of (\ref{PB160}). This exercise clearly shows an intimate connection between the $SD_{B\wedge F}$ with a gauge invariant version through a dual transformation.  However, although establishing the dual connection, the result obtained in (\ref{PB160}) produces a set of field equations involving higher derivatives that will produce more solutions than the original set.  Besides they are not the usual $TM_{B\wedge F}$ model.  In the next section we shall discuss a dynamical gauge embedding procedure that will clearly produce an equivalent gauge invariant model.

\section{The Gauge Invariant $B\wedge F$ Theory}

In previous works \cite{IW,AINRW} we have used the dynamical gauge embedding formalism to study dual equivalence in (2+1)
dimensions in diverse situations with models involving the presence of the topological Chern-Simons term. In this
section we extend that technique to study duality symmetry among four dimensional models involving the presence of a topological $B\wedge F$ term.

Our basic goal is to transform the symmetry (\ref{PB140}) that is hidden in the Lagrangian (\ref{PB10})
into a local gauge symmetry by lifting the global parameter $\Lambda$ into its local form, i.e.,
$\Lambda\to \Lambda(x^\mu)$.
The method works by looking for an (weakly) equivalent description of the original
theory which may be obtained by adding a function $f(K_\mu , M_{\mu\nu})$ to the Lagrangian
(\ref{PB10}). Here $K_{\mu}$ and $M_{\mu\nu}$ are the Euler tensors, defined by the variation
\begin{eqnarray}
\label{PB180}
\delta {\cal L}_{SD}^{(0)} &=& K_\mu \delta A^\mu + M_{\mu\nu} \delta B^{\mu\nu}
\end{eqnarray}
whose kernels give the equations of motion for the $A_\mu$ and $B_{\mu\nu}$ fields, respectively.  The minimal requirement for $f(K_\mu , M_{\mu\nu})$ is that it must be chosen such that it vanishes on the space of solutions of (\ref{PB10}), viz. $f(0,0)=0$, so that the effective Lagrangian ${\cal L}_{eff}$
\begin{equation}
\label{PB190}
{\cal L}_{SD}^{(0)} \to {\cal L}_{eff}= {\cal L}_{SD}^{(0)} + f(K_\mu , M_{\mu\nu})
\end{equation}
is dynamically equivalent to ${\cal L}_{SD}^{(0)}$. To find the specific form of this function that also induces
a gauge symmetry into ${\cal L}_{SD}^{(0)}$ we work iteratively.
To this end we compute the variations (\ref{PB180}) of ${\cal L}_{SD}^{(0)}$ to find the Euler
tensors as
\begin{eqnarray}
\label{PB200}
K_\mu  &=& m^2 A_\mu - \frac {\chi\, \theta}2 \epsilon_{\mu\nu\rho\lambda} \partial^\nu B^{\rho\lambda}   \nonumber\\
M_{\mu\nu} &=&  - \frac 12 B_{\mu\nu} + \frac {\chi\, \theta}2 \epsilon_{\mu\nu\lambda\rho}\partial^\lambda A^\rho
\end{eqnarray}
and define the first-iterated Lagrangian as,
\begin{equation}
\label{PB210}
{\cal L}_{SD}^{(1)} = {\cal L}_{SD}^{(0)} - a_\mu K^\mu\, - b_{\mu\nu} M^{\mu\nu}
\end{equation}
with the Euler tensors being imposed as constraints and the new fields, $a_\mu$ and $b_{\mu\nu}$, to be identified with ancillary gauge fields, acting as a Lagrange multipliers.

The transformation properties of the auxiliary fields $a_\mu$ and $b_{\mu\nu}$ accompanying the basic field
transformations (\ref{PB140}) is chosen so as to cancel the variation of
${\cal L}_{SD}^{(0)}$, which gives
\begin{eqnarray}
\label{PB220}
\delta a_\mu &=& \delta A_\mu \nonumber\\
\delta b_{\mu\nu} &=& \delta B_{\mu\nu} 
\end{eqnarray}
A simple algebra then shows
\begin{eqnarray}
\label{PB230}
\delta {\cal L}_{SD}^{(1)} &=&  - a_\mu \,\delta K^\mu - b_{\mu\nu} \,\delta M^{\mu\nu}\nonumber\\
&=& \delta \left( - \frac{m^2}2 a_\mu a^\mu + \frac 14 b_{\mu\nu}b^{\mu\nu}\right)
\end{eqnarray}
where we have used (\ref{PB140}) and (\ref{PB220}).  Because of (\ref{PB230}), the
second iterated Lagrangian is unambiguously defined as
\begin{equation}
\label{PB240}
{\cal L}_{SD}^{(2)} = {\cal L}_{SD}^{(1)} + \frac{m^2}2 a_\mu a^\mu - \frac 14 b_{\mu\nu}b^{\mu\nu}
\end{equation}
that is automatically gauge invariant under the combined local transformation of the original set of fields
($A_\mu$, $B_{\mu\nu}$) and the auxiliary fields ($a_\mu$, $b_{\mu\nu}$).

We have therefore succeed in transforming the global $SD_{B\wedge F}$ theory into a locally
invariant gauge theory.  We may now take advantage of the Gaussian
character displayed by the auxiliary field to rewrite (\ref{PB240})
as an effective action depending only on the original variables ($A_\mu$, $B_{\mu\nu}$).
To this end we use (\ref{PB240}) to solve for the fields $a_\mu$ and $b_{\mu\nu}$ (call the
solutions $\bar a_\mu$ and $\bar b_{\mu\nu}$ collectively by $\bar h_{\{\mu\}}$), and replace it back into (\ref{PB240}) to find
\begin{eqnarray}
\label{PB100}
{\cal L}_{eff}&=&{\cal L}_{SD}^{(2)}\mid_{h_{\{\mu\}} = \bar h_{\{\mu\}}} \nonumber\\
&=& {\cal L}_{SD}^{(0)} - \frac 1{2m^2} K_\mu K^\mu  + M_{\mu\nu} M^{\mu\nu} 
\end{eqnarray}
from which we identify the function $f(K_\mu , M_{\mu\nu})$ in (\ref{PB190}).
This dynamically modified action can be rewritten to give the $TM_{B\wedge F}$ theory,
\begin{equation}
\label{BI40}
{\cal L}_{eff} =  \frac 1{12 m^2} \:H_{\mu\nu\lambda} H^{\mu\nu\lambda}  - \frac 14\: F_{\mu\nu} F^{\mu\nu}  - \frac \chi{2\,\theta} \:\epsilon^{\mu\nu\alpha\beta}\: A_\mu \partial_\nu B_{\alpha\beta}\; .
\end{equation}
after the scaling $\theta\, A_\mu\to A_\mu$ and $\theta\, B_{\mu\nu}\to B_{\mu\nu}$ is performed. Notice the inversion of the coupling constant $\theta\to 1/\theta$ resulting from the duality mapping.
It becomes clear from the above derivation that the difference between these
two models is given by a function of the Euler tensors of the $SD_{B\wedge F}$ model that
vanishes over its space of solutions.  This establishes the
dynamical equivalence between the $SD_{B\wedge F}$ and the $TM_{B\wedge F}$ theory.

\section{The minimal coupling with fermionic matter}

Once the duality mapping between the free theories has been established one is ready to consider the requirements for the existence of duality when the coupling with dynamical matter is included.  In this section we consider the case of U(1) charged fermionic matter. For clarity, we consider first the situation where only the vector field, in the self-dual representation, is minimally coupled to the fermionic current. This seems appropriate since it illustrates the main features of the duality via gauge embedding.  The results for the full coupling are then quickly presented in the following subsection.

\subsection{Single Coupling}
Let us consider first a Dirac field minimally coupled to the vector field $A_\mu$ specified
by the $SD_{B\wedge F}$ model. To this end let us introduce the fermionic current
\begin{eqnarray}
\label{100}
J^\mu = \bar{\psi}\gamma^\mu \psi
\end{eqnarray}
and define the rank-2 charge $J^{\mu\nu}$ as the (derivative) dual of the fermionic current $J_\mu$,
\begin{equation}
J^{\mu\nu} = - \frac \theta{2 m^2} \epsilon^{\mu\nu\alpha\beta}\partial_\alpha J_\beta
\end{equation}
so that the interacting Lagrangian becomes
\begin{equation}
\label{40}
{\cal L}_{min}^{(0)} = {\cal L}_{SD}^{(0)}  - e A_\mu J^\mu + 
{\cal L}_{D}  \; , 
\end{equation}
where $M$ is the fermion mass. Here the Dirac Lagrangian is,
\begin{equation}
\label{45}
{\cal L}_{D} =  \bar{\psi}(i\partial\!\!\! /  -M)\psi \; .
\end{equation}
The fermionic field is treated as an spectator in the dual transformation between the gauge fields from the SD to the TM sectors. But to remain as a bystander field the coupling to the gauge fields and to itself has to be readjusted in the TM, as shown below.

As before, our basic strategy is to transform the hidden symmetry of the Lagrangian (\ref{40})
into a local gauge symmetry.
A variation of the Lagrangian (\ref{40}) gives the Euler vectors as,
\begin{eqnarray}
\label{50}
K_\mu \to K_\mu^D &=& K_\mu - e J_\mu\nonumber\\
M_{\mu\nu} \to M_{\mu\nu}^D &=& M_{\mu\nu} \, .
\end{eqnarray}
>From now on we follow the same steps as in the free case just making the replacement $K_\mu \to K_\mu^D $ and 
$M_{\mu\nu} \to M_{\mu\nu}^D$ to obtain an effective action as,
\begin{eqnarray}
\label{1001}
{\cal L}_{eff} = {\cal L}_{min}^{(0)} - \frac 1{2m^2} K_\mu^D K^{D\,\mu}  - M_{\mu\nu}^D M^{D\,\mu\nu}
\end{eqnarray}
A further manipulation shows the presence of a new term, compared to the free case and of a non-minimal coupling
\begin{eqnarray}
\label{PB1101}
{\cal L}_{eff} &=&  \frac 1{12 m^2} \:H_{\mu\nu\lambda} H^{\mu\nu\lambda}  - \frac 14\: F_{\mu\nu} F^{\mu\nu}  - \frac \chi{2\,\theta} \:\epsilon^{\mu\nu\alpha\beta}\: A_\mu\partial_\nu B_{\alpha\beta}\; .
\nonumber\\
&+& 
\bar{\psi}(i\partial\!\!\! /  -M)\psi - \frac {e^2}{2 m^2} J_\mu J^\mu
- \frac{\chi}{\theta}\, e\, J^{\mu\nu} B_{\mu\nu}
\end{eqnarray}
The presence of these two terms involving the matter current are worth discussing since they play important role in preserving the structure of the fermionic sector upon application of the dual mapping upon the gauge sector. They are
the Thirring-like self-interaction and the non-minimal magnetic-like interaction of the dual of the fermionic current with $B_{\mu\nu}$.  It is interesting to observe that the minimal coupling involving the $A_\mu$ field became, through the dualization process, a non-minimal coupling for the $B_{\mu\nu}$ field.  This swapping of the coupling is hardly a surprise. It is the manifestation, in the latent sector of duality, of the traditional duality inversion and, as far as we know, this phenomenon has not been reported before. Notice that both terms appear naturally as a consequence of the
embedding algorithm.  The presence of these terms, as we show in the sequel, are important to maintain the dynamical structure of the fermionic matter in both representations of the dual pair \cite{GMdS}. This can be seen by carefully
examining the dynamics of the fermionic sector for both theories. 
To see this we compute the fermionic equation for the $SD_{B\wedge F}$ sector, that reads
\begin{eqnarray}
\label{AIS05234}
\left(i\,\slash\!\!\!{\partial} - M\right)\psi =
e A^\mu \,\gamma_{\mu}\,\psi\,.
\end{eqnarray}
To eliminate the bosonic field $A^\mu$ in favor of the fermionic one we rewrite the equations
of motion for the $SD_{B\wedge F}$ sector as,
\begin{eqnarray}
\label{30}
A^\mu &=& \frac {\chi\theta}{2 m^2} \epsilon^{\mu\nu\lambda\rho}\partial_\nu B_{\lambda\rho} + e J^\mu \\
B^{\mu\nu} &=&  \chi\theta \:\epsilon^{\mu\nu\lambda\rho}\partial_\lambda  A_{\rho}
\end{eqnarray}
which can be combined to give, after using the constraint (\ref{PB40}),
\begin{equation}
\label{33}
R^{-1} A_\mu = e J_\mu
\end{equation}
where $R$ is a differential operator such that its inverse is the wave-operator,
defined as
\begin{eqnarray}
\label{AIS08}
R^{-1} = \partial_{\lambda}\partial^{\lambda} +\frac {m^{2}}{\theta^2}\,.
\end{eqnarray}
Substituting this equation back in the matter equation gives,
\begin{eqnarray}
\label{AIS05235}
\left(i\,\slash\!\!\!{\partial} - M\right)\psi = 
e^2 R \, J^\mu\,\gamma_{\mu}\,\psi\,,
\end{eqnarray}
that is a nonlinear, integro-differential equation, now written completely in terms of the fermionic fields.

Next we compare this result with the equations of motion
for the fermionic matter from the $TM_{B\wedge F}$ sector, that reads
\begin{eqnarray}
\label{AIS0523}
\left(i\,\slash\!\!\!{\partial} - M\right)\psi = \, \chi \, e\,\, \mbox{}^*A^\mu \gamma_\mu \, \psi
+ \frac {e^2}{m^2} J^\mu \gamma_\mu \, \psi
\end{eqnarray}
where we have to obtain the bosonic functional $\mbox{}^*A^\mu \to \mbox{}^*A^\mu\left(\psi\right)$
in terms of the fermions from the equations of motion for the gauge fields in the $TM_{B\wedge F}$ sector,
\begin{eqnarray}
\label{37}
\mbox{}^*A^\mu &=& -\frac {\chi\theta}{2 m^2} \epsilon^{\mu\nu\lambda\rho}\partial_\nu \mbox{}^*B_{\lambda\rho}\nonumber \\
\mbox{}^*B^{\mu\nu} &=& \, \chi\theta \:\epsilon^{\mu\nu\lambda\rho}\partial_\lambda  \mbox{}^*A_{\rho} - 2\, e \,
G^{\mu\nu}
\end{eqnarray}
that can be combined to give,
\begin{eqnarray}
\label{39}
\mbox{}^*A^\mu\left(\psi\right) = \chi \, e \left( \frac {\theta^2}{m^2}  - R \right)\, J^\mu\, .
\end{eqnarray}
When inserted into (\ref{AIS0523}) gives
\begin{eqnarray}
\label{41}
\left(i\,\slash\!\!\!{\partial} - M\right)\psi = 
e^2 R \, J^\mu\,\gamma_{\mu}\,\psi\,,
\end{eqnarray}
which agrees with the dynamical equation for the fermions in the SD side. 
We just stress the importance of the Thirring like self-interaction for the fermions and the non-minimal interaction with the tensor field to keep the dynamics of the latent, fermionic sector unaltered.

\subsection{Double Coupling}

It is now a simple task to consider the full coupling of the bosonic fields $A_\mu$ and $B_{\mu\nu}$ to fermionic matter.
To this end we introduce the rank-2 current
\begin{eqnarray}
\label{399}
G^{\mu\nu} = {\cal C} \bar\psi \gamma^\mu\gamma^\nu\psi
\end{eqnarray}
where ${\cal C}$ is a complex normalization constant and its dual
\begin{eqnarray}
\label{3991}
G^{\mu} = \theta \, \epsilon^{\mu\lambda\rho\sigma}\partial_\lambda G_{\rho\sigma}
\end{eqnarray}
The interacting Lagrangian now takes the form
\begin{equation}
\label{401}
{\cal L}_{min}^{(0)} = {\cal L}_{SD}^{(0)}  - e A_\mu J^\mu +  g B_{\mu\nu}G^{\mu\nu} +
{\cal L}_{D}  \; , 
\end{equation}
with $e$ and $g$ being the strengths of the coupling with $A_\mu$ and $B_{\mu\nu}$, respectively.
The effective, gauge invariant action is obtained directly from (\ref{PB100}) just operating the replacement
\begin{eqnarray}
\label{501}
K_\mu \to K_\mu^D &=&  K_\mu - e J_\mu \nonumber\\
M_{\mu\nu} \to M_{\mu\nu}^D &=& M_{\mu\nu} + g  G_{\mu\nu}
\end{eqnarray}
to produce
\begin{eqnarray}
\label{1000}
{\cal L}_{eff} &=& {\cal L}_{min}^{(0)} - \frac 1{2m^2} K_\mu^D K^{D\,\mu}  - M_{\mu\nu}^D M^{D\,\mu\nu} 
\end{eqnarray}
which, after some algebraic manipulation, gives
\begin{eqnarray}
\label{PB1102}
{\cal L}_{eff} &=&  {\cal L}_D + \frac 1{12 m^2} \:H_{\mu\nu\lambda} H^{\mu\nu\lambda}  - \frac 14\: F_{\mu\nu} F^{\mu\nu}  - \frac \chi{2\,\theta} \:\epsilon^{\mu\nu\alpha\beta}\: A_\mu\partial_\nu B_{\alpha\beta} \nonumber\\
&-&   \frac {e^2}{2 m^2}\, J_\mu J^\mu + g^2 \, G^{\mu\nu}G_{\mu\nu}
- \frac{\chi\, e}{\theta}\,  J^{\mu\nu} B_{\mu\nu}  +\frac {\chi\,g}\theta A_\mu\, G^{\mu}
\end{eqnarray}
{}From this result it becomes clear the full action of the dual mapping over the active and passive fields involved in the transformation. Notice the exchange of the minimal coupling adopted in the SD sector into a non minimal, magnetic like interaction in the TM sector, including a swapping between the fields and currents and the presence of the current-current interaction for the fermionic sector which is needed to maintain the dynamics of the spectator field unmodified. this is easily checked by just computing the equations of motion for the Dirac fields in both representations to obtain,
\begin{eqnarray}
\label{410}
\left(i\,\slash\!\!\!{\partial} - M\right)\psi &=& \frac R{\theta^2}\gamma_\mu\left[\left(  {e^2} \, J^\mu
- {\chi ge}\, G^\mu\right) + {2\, m^2{\cal C}}\left(\chi ge  J^{\mu\nu} + g^2  G^{\mu\nu}\right)
\gamma_\nu\right]\psi
\end{eqnarray}

\section{Conclusions}

In this work we studied dual equivalence in four dimensional topological models, namely, between the $B\wedge F$ self-dual ($SD_{B\wedge F}$) and the $B\wedge F$ topologically massive ($TM_{B\wedge F}$) models using an iterative procedure of gauge embedding that produces the dual mapping. We defined a new derivative type of duality mapping, very much like the one adoted in the three-dimensional case and proved the self and antiself-duality property of the $SD_{B\wedge F}$ model, according to the relative sign of the topological term.
Working out the free case firstly, where the A and B fields participate actively in the dual transformation we obseved, as expected, the traditional inversion in the coupling constant. The coupling to dynamical fermionic matter, which acts as a spectator field in the dual transformation, brought into the scene some new features. The apearance of a Thirring like self-interaction term in the dualized theory, that had already been observed in the (2+1) case, as well as the shift from minimal to non minimal coupling. However, in this case we observed a swapping of the couplings from a tensor to another. This is a new result due to the presence of tensors of distinct ranks participating actively in the dual transformation.  We proved that the presence of these terms are demanded to maintain the equivalent dynamics in the fermionic sector in either representations of the duality. The cases where the active tensors appear non linearly and the coupling with bosonic matter are postponed for a forthcoming publication.

\vspace{.5cm}

\noindent ACKNOWLEDGMENTS: This work is partially supported by CNPq, CAPES,
FAPERJ and FUJB, Brazilian Research Agencies.  CW thanks the Physics
Department of UFPB for the kind hospitality during the course of this
investigation.

\end{document}